\documentclass[12pt,a4paper]{article}
\usepackage{a4wide}
\usepackage{amsmath}
\usepackage{amssymb}
\usepackage{epsfig}
\usepackage{subfigure}
\usepackage{exscale}
\usepackage{float}
\usepackage{bbm}
\usepackage[numbers,sort&compress]{natbib}
\usepackage{pst-plot, pstricks,pst-math}
\usepackage{fancybox,amssymb,color}
\usepackage{graphicx}
\usepackage{pstricks, color, graphicx, epsfig, psfrag}
\usepackage{amsfonts,amsmath,amssymb,slashed}
\usepackage{dsfont}
\usepackage{bbm,bm}
\usepackage{fancyhdr, a4wide}
\usepackage[english]{babel}
\usepackage{subfigure}

\makeatletter
\@addtoreset{equation}{section}
\makeatother

\title{Antiferromagnetic Dispersion Relations and Nature of Magnon Pressure}

\author{Christoph P.\ Hofmann$^a$ \\ \\
\normalsize{$^a$ Facultad de Ciencias, Universidad de Colima} \\
\vspace{0.3cm}
\normalsize{Bernal D\'iaz del Castillo 340, Colima C.P.\ 28045, Mexico} \\}

\begin{document}
\maketitle

\begin{abstract} \normalsize

We derive higher-order corrections in the magnon dispersion relations for two- and three-dimensional antiferromagnets exposed to magnetic
and staggered fields that are mutually aligned. "Dressing" the magnons is the prerequisite to separate the low-temperature representation
of the pressure into a piece due to noninteracting magnons and a piece that corresponds to the magnon-magnon interaction. Both in two and
three spatial dimensions, the interaction in the pressure turns out to be attractive. While concrete figures refer to the
spin-$\frac{1}{2}$ square-lattice and the spin-$\frac{1}{2}$ simple cubic lattice antiferromagnet, our results are valid for arbitrary
bipartite geometry.

\end{abstract}

\maketitle

\section{Introduction}
\label{Intro}

The impact of external magnetic fields on antiferromagnetic systems -- both in two and three spatial dimensions -- has been studied by
various authors employing different techniques: (modified) spin-wave theory
\citep{Kub52,Ogu60,Joe62,Tak89,HWA92,AS93,ZN98a,ZN98b,SSKPWLB05,HSSK07,KSHK08},
Green's functions
\citep{Mor72,Gho73,LR74,LR75,WHG14},
series expansions
\citep{HWA91,Pan98,Pan99,Pan00},
Monte Carlo simulations
\citep{San99,SS02,PR15},
exact diagonalization
\citep{FKLM92,LL09},
and yet other methods
\citep{Fal64,PSS69,CS70a,MM79,BFD90,Glu93,SSS94,BS04,FS04,NS07,VK09,NVSFS18}.

Still, the concrete configuration of antiferromagnets exposed to mutually aligned magnetic and staggered fields, has not been studied in a
fully systematic way in the aforementioned references. In particular, higher-order effects where the spin-wave interaction becomes
relevant, have been neglected. It is the goal of the present investigation to help to close this gap. Our approach is based on magnon
effective field theory that has been established in earlier work -- see
Refs.~\citep{HH69,CHN89,NZ89,Fis89,HL90,HN93,Leu94a}
-- and has specifically been applied to two- and three-dimensional antiferromagnets in
Refs.~\citep{Hof99a,Hof99b,RS99a,RS99b,RS00,Hof10,Hof17,Hof17b,BH17,BH20,Hof20a,Hof20b,Hof20c,Hof20d}.

Here we first calculate the two-point functions for antiferromagnetic magnons residing in antiferromagnets exposed to mutually aligned
magnetic and staggered fields. This enables us to derive the corresponding dispersion relations, in particular to evaluate higher-order
corrections. Within this dressed magnon framework we can isolate in the pressure the piece that is due to noninteracting (but dressed)
magnons, and are then left with the piece that can be attributed to the genuine magnon-magnon interaction that emerges at two-loop order in
the systematic effective field theory calculation.

We find that the spin-wave interaction in the pressure is attractive in the entire parameter region of external magnetic and staggered
fields we are exploring. If the magnetic field is turned off, the interaction tends to zero and the system is described by the
noninteracting magnon gas. An important observation -- both in two and three spatial dimensions -- is that the contribution due to the
magnon-magnon interaction in the pressure does not involve any next-to-leading order (NLO) effective constants, but uniquely depends on the
two leading order effective constants that are well-known: spin stiffness and order parameter, i.e., the staggered magnetization at zero
temperature. As it turns out, the interaction in the case of three-dimensional antiferromagnets is quite weak.

In concrete figures we resort to the spin-$\frac{1}{2}$ square-lattice and the spin-$\frac{1}{2}$ simple cubic lattice antiferromagnet --
where the numerical values for spin stiffness and order parameter are available. It should be stressed, however, that our two-loop
representations for the pressure are also fully rigorous and predictive for any other bipartite lattice. Most importantly, the observation
that the nature of the interaction in the pressure is attractive is valid for any such system.

The article is organized as follows. In Sec.~\ref{fedDressed}, after a few general comments on antiferromagnets in magnetic fields aligned
with the order parameter, we derive the two-point functions and the dispersion laws for the magnons up to next-to-leading order in the
effective expansion for two- and three-dimensional antiferromagnets. We then isolate the genuine spin-wave interaction piece in the free
energy density. In Sec.~\ref{pressure} we discuss the low-temperature representation of the pressure and show that the spin-wave
interaction is attractive in presence of magnetic and staggered fields. In plots we refer to spin-$\frac{1}{2}$ square-lattice and
spin-$\frac{1}{2}$ simple cubic lattice antiferromagnets. In Sec.~\ref{conclusions} we finally conclude.

\section{Dispersion Relations and Dressed Magnons}
\label{fedDressed}

\subsection{Preliminaries}
\label{prelim}

The underlying model that describes antiferromagnetic systems is the isotropic Heisenberg Hamiltonian augmented by an external magnetic
(${\vec H}$) and a staggered (${\vec H_s}$) field,
\begin{equation}
\label{HeisenbergZeemanH}
{\cal H} \, = \, - J \, \sum_{n.n.} {\vec S}_m \! \cdot {\vec S}_n \, - \, \sum_n {\vec S}_n \cdot {\vec H} \, - \, \sum_n (-1)^n {\vec S}_n
\! \cdot {\vec H_s} \, , \qquad J < 0 \, , \quad J = \mbox{const.} \, ,
\end{equation}
where "n.n." means we are summing over nearest neighbor spins only. The lattice is furthermore assumed to be bipartite.

In the present analysis we consider the configuration of mutually parallel magnetic and staggered fields,
\begin{equation}
\label{externalFields}
{\vec H} = (H,0,0) \, , \qquad {\vec H}_s = (H_s,0,0) \, , \qquad H, H_s > 0 \, ,
\end{equation}
that are furthermore aligned with the staggered magnetization at zero temperature which represents the order parameter. The magnon
dispersion laws\footnote{Note that the spin-wave velocity $v$ does not appear as we have set it to one.} are then characterized by an
energy gap and read (see Refs.~\citep{ABK61,Nol86,Hof20a})
\begin{eqnarray}
\label{disprelAFHparallel}
\omega_{+} & = & \sqrt{{\vec p \,}^2 + \frac{M_s H_s}{\rho_s}} + H \, , \nonumber \\
\omega_{-} & = & \sqrt{{\vec p \,}^2 + \frac{M_s H_s}{\rho_s}} - H \, .
\end{eqnarray}
The quantities $\rho_s$ and $M_s$ are the spin stiffness and the staggered magnetization at $T$=0, respectively. Within the effective field
theory perspective these constitute the two so-called leading order effective constants.

While the above dispersion relations -- that apply to two- and three-dimensional antiferromagnets alike -- only involve $\rho_s$ and $M_s$,
this is no longer the case at subleading orders. As it turns out, the dispersion law for three-dimensional antiferromagnets, in addition,
involves next-to-leading order effective constants.

It is important to point out that the spin-wave branch $\omega_{-}$ becomes negative, unless the criterion
\begin{equation}
\label{stabilityCondition}
H_s > \frac{\rho_s}{M_s} \, H^2
\end{equation}
is satisfied. Here we take it for granted that this stability criterion is met. Otherwise the order parameter changes its orientation and
an alternative ground-state configuration is realized where the magnetic field is perpendicular to the staggered magnetization -- within
effective field theory this case has been investigated in Refs.~\citep{Hof17,BH17,BH20,Hof20b}.

One of our objectives is to discuss the magnon pressure, in particular to determine whether the magnon-magnon interaction in the pressure
is attractive or repulsive, and to explore how the interaction is affected by temperature, magnetic and staggered field. To this end -- as
will become clear below -- we have to calculate the two-point functions for the two types of magnons and evaluate their dispersion laws to
higher orders. It should be noted that the organization of Feynman diagrams in the effective low-energy expansion depends on the space-time
dimension: in two (three) spatial dimensions each additional magnon loop corresponds to a suppression by one (two) powers of energy or
temperature.\footnote{See, e.g., Sec.~III of Ref.~\citep{Hof10}.} Therefore we have to address two- and three-dimensional antiferromagnets
separately.

\subsection{Two-Dimensional Antiferromagnets}
\label{dressed2d}

\begin{figure}
\begin{center}
\hbox{
\includegraphics[width=12.6cm]{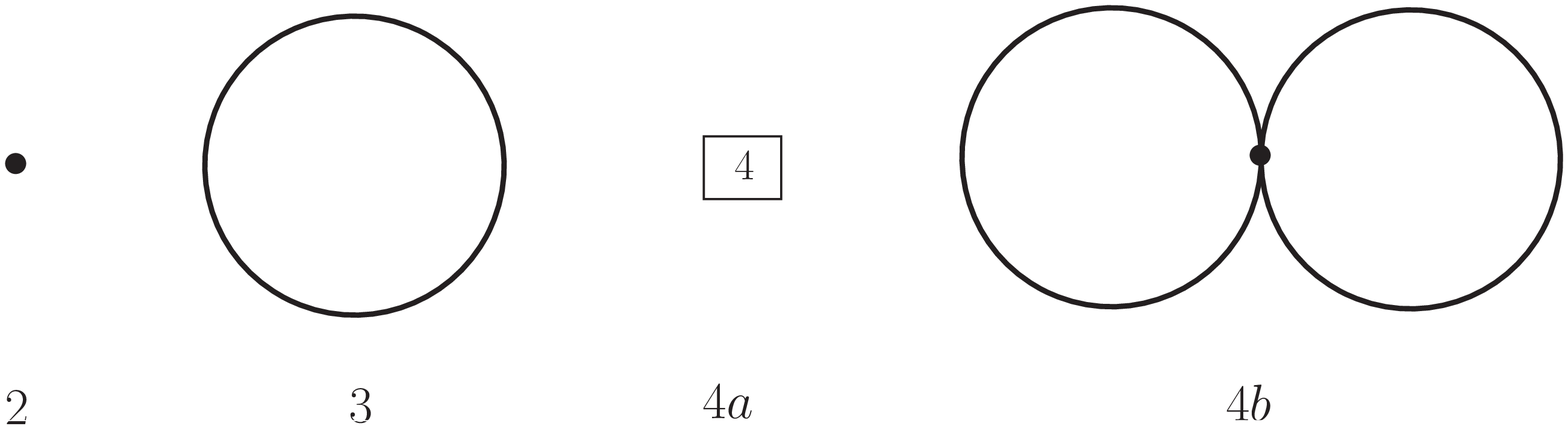}}
\vspace{4mm}
\hbox{
\includegraphics[width=8.1cm]{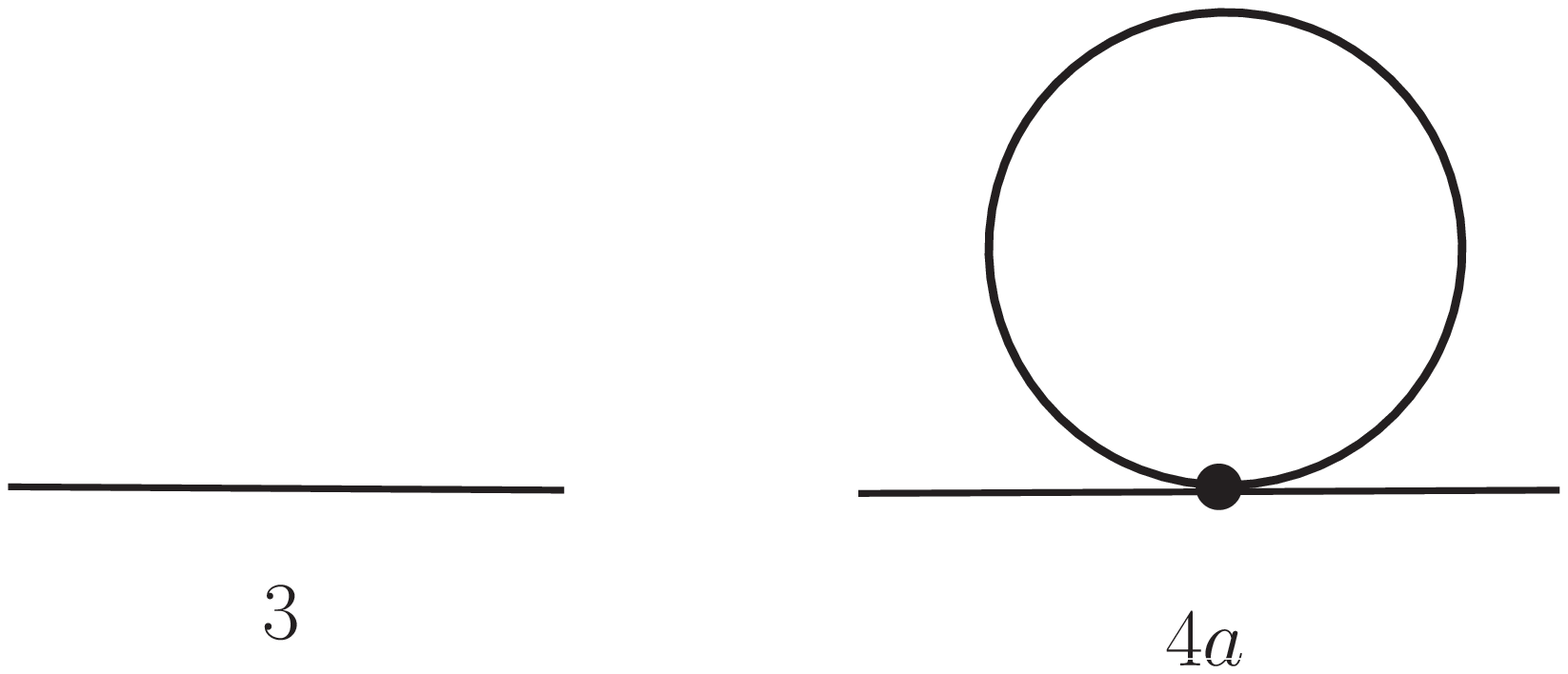}}
\end{center}
\caption{Two-dimensional antiferromagnets in mutually aligned magnetic and staggered fields. Upper panel: Partition function diagrams up to
two-loop order. Lower panel: Two-point function diagrams up to one-loop order. Filled circles constitute vertices from the leading order
effective Lagrangian, while the "4" in the box represents the NLO effective Lagrangian.}
\label{figure1}
\end{figure}

Let us first consider antiferromagnetic films. The diagrams for the free energy density up to two-loop order are depicted in the upper
panel of Fig.~\ref{figure1}.\footnote{Details on the effective loop evaluation can be found in Ref.~\citep{Hof20a}.} The leading
finite-temperature diagram $3$ (order $T^3$) merely involves noninteracting magnons. The interaction starts manifesting itself through the
finite-temperature two-loop diagram $4b$ (order $T^4$). The essential point is to realize that the two-loop diagram $4b$ -- apart from
describing the leading magnon-magnon interaction contribution at finite temperature -- also contains a piece that refers to the
magnon-magnon interaction at zero temperature. This $T$=0 piece modifies the  magnon dispersion relations, i.e., it "dresses" the magnons.
In order to hence have a clear definition of the interaction at {\it finite} temperature, the free energy density has to be expressed in
terms of these dressed magnons.

Following this strategy we now derive the two-point function for the two antiferromagnetic magnons in external fields and extract their
dispersion relation. Evaluating the magnon two-point function $\tau_{\pm}(x-y)$ up to one-loop order is straightforward, because there are
only two diagrams contributing, as shown in the lower panel of Fig.~\ref{figure1}. The leading contribution (diagram 3), yields the $T$=0
propagator $\Delta_{\pm}(x-y)$ for magnon $+$ and magnon $-$,
\begin{equation}
\tau_{\pm}^3(x-y) = \Delta_{\pm}(x-y) = \int \!\! \frac{\mbox{d}^d \! p}{{(2 \pi)}^d} \,
\frac{e^{ip(x-y)}}{p_4^2 + {\vec p \,}^2 + M^2 \pm 2 i H p_4 - H^2} \, ,
\end{equation}
regularized in the space-time dimension $d$. The magnon "mass" $M$ is defined as
\begin{equation}
M^2 = \frac{M_s H_s}{\rho_s} \, .
\end{equation}
The only correction comes from one-loop graph 4a and reads
\begin{equation}
\label{TwoPointA}
\tau_{\pm}^{4a}(x-y) = \pm \frac{2 i H}{\rho_s} \, \Delta(0) \, \int \!\! \frac{\mbox{d}^d \! p}{{(2 \pi)}^d}
\, \frac{p_4 \, e^{ip(x-y)}}{{(p_4^2 + {\vec p \,}^2 + M^2 \pm 2 i H p_4 - H^2)}^2} \, ,
\end{equation}
and can be embedded into the physical two-point function $\tau_{\pm}(x-y)$ via
\begin{eqnarray}
\tau_{\pm}(x-y) & = & \int \!\! \frac{\mbox{d}^d \! p}{{(2 \pi)}^d} \,
\frac{e^{ip(x-y)}}{p_4^2 + {\vec p \,}^2 + M^2  \pm 2 i H p_4 - H^2 + X_{\pm}} \nonumber \\
& = & \int \!\! \frac{\mbox{d}^d \! p}{{(2 \pi)}^d} \, \frac{e^{ip(x-y)}}{p_4^2 + {\vec p \,}^2 + M^2 \pm 2 i H p_4 - H^2} \nonumber \\
& & \times \Bigg\{ 1 - \frac{X_{\pm}}{p_4^2 + {\vec p \,}^2 + M^2 \pm 2 i H p_4 - H^2} + {\cal O}(X^2/{\cal D}^2) \Bigg\} \, .
\end{eqnarray}
The quantity $\Delta(0)$ in Eq.~(\ref{TwoPointA}) is the dimensionally regularized zero-temperature magnon propagator at the coordinate
origin $x$=0 when no magnetic field is present,
\begin{equation}
\label{regprop}
\Delta(0) = \int \frac{{\mbox{d}}^d p}{{(2 \pi)}^d} \, \frac{1}{p^2 + M^2}
= {\int}_{\!\!\!0}^{\infty} \mbox{d} \rho \, (4 \pi \rho)^{-d/2} e^{-\rho M^2} \, .
\end{equation}
While the dispersion relation
\begin{equation}
{\cal D} = p_4^2 + {\vec p \,}^2 + M^2 \pm 2 i H p_4 - H^2
\end{equation}
is tied to the leading order propagator $\Delta_{\pm}(x-y)$, the correction $X_{\pm}$ takes into account the next-to-leading order
contribution $\tau_{\pm}^{4a}(x-y)$.

The physical limit $d \to 3$ can be taken without problems: the quantity $\Delta(0)$ is not singular in two spatial dimensions, but
remains finite when the regularization is removed,
\begin{equation}
\lim_{d \to 3} \Delta(0) = - \frac{M}{4 \pi} = - \frac{\sqrt{M_s H_s}}{4 \pi \sqrt{\rho_s}} \, ,
\end{equation}
such that the dispersion relations for the "dressed" magnons take the form 
\begin{equation}
\omega_{\pm} = \sqrt{{\vec p \,}^2 + \frac{M_s H_s}{\rho_s} + \frac{H^2 \sqrt{M_s H_s}}{2 \pi \rho^{\frac{3}{2}}_s}} \pm H
\pm \frac{H \sqrt{M_s H_s}}{4 \pi \rho^{\frac{3}{2}}_s} \, .
\end{equation}
On the basis of these relations we can now determine the portion in the free energy density that is associated with noninteracting magnons
by means of
\begin{equation}
\label{freeEnergyBasic2d}
z^{free} = z_0^{free} + \frac{T}{{(2 \pi)}^2} \, \int \!\! \mbox{d}^2 \! p \, \ln \Big[ 1 - e^{- \omega_{+}({\vec p}) / T} \Big]
+ \frac{T}{{(2 \pi)}^2} \, \int \!\! \mbox{d}^2 \! p \, \ln \Big[ 1 - e^{- \omega_{-}({\vec p}) / T} \Big] \, ,
\end{equation}
where $z_0^{free}$ is the vacuum energy density of the noninteracting magnons. The corrections to the leading order dispersion law appear as
\begin{equation}
\omega_{\pm}({\vec p}) = \sqrt{{\vec p \,}^2 + \frac{M_s H_s}{\rho_s} + \epsilon^A_{\pm}} \pm H + \epsilon^B_{\pm} \, .
\end{equation}
We thus consider the pertinent expansions
\begin{eqnarray}
\exp \! \Big(- \frac{\omega_{\pm}}{T} \Big)
& \approx & \exp \! \Big( - \frac{\omega_0 \pm H}{T} \Big) \Bigg\{ 1 - \frac{\epsilon^B_{\pm}}{T} - \frac{\epsilon^A_{\pm}}{2 \omega_0 T}
+ {\cal O}(\epsilon^2) \Bigg\} \, , \nonumber \\
& & \omega_0 = \sqrt{{\vec p \,}^2 + \frac{M_s H_s}{\rho_s}} \, ,
\end{eqnarray}
and
\begin{eqnarray}
\ln \! \Big( 1 - e^{-\frac{\omega_{\pm}}{T}} \Big)
& \approx & \ln \! \Big( 1 - e^{- (\omega_0 \pm H)/ T} \Big) \nonumber \\
& & + \frac{1}{T} \, \Bigg\{ \epsilon^B_{\pm} + \frac{\epsilon^A_{\pm}}{2 \omega_0} \Bigg\} \, \frac{1}{e^{(\omega_0 \pm H)/ T}-1} \, ,
\end{eqnarray}
integrate over momentum according to Eq.~(\ref{freeEnergyBasic2d}), and end up with the portion in the free energy density that is due to
noninteracting magnons:
\begin{equation}
\label{freeDressedED}
z^{free}= - {\hat h}_0 \, T^3
- \frac{\sqrt{M_s H_s} H}{4 \pi \rho_s^{3/2}} \, \frac{\partial {\hat h}_0}{\partial H} \, T^3
+ \frac{\sqrt{M_s H_s} H^2}{2 \pi \rho_s^{3/2}} \, {\hat h}_1 \, T + z^{free}_0 \, .
\end{equation}
The kinematical functions ${\hat h}_0$ (or equivalently: ${\hat g}_0)$ and ${\hat h}_1$ (or equivalently: ${\hat g}_1)$ are  
\begin{eqnarray}
\label{defh0}
{\hat h}_0 & = & \frac{{\hat g}_0}{T^3} \\
& = & -\frac{1}{2 \pi T^2} \int_0^{\infty} \!\! \mbox{d}p \, p \, \Bigg\{ \ln \Big[ 1 - e^{- (\sqrt{p^2 + M^2} + H)/T} \Big] +
\ln \Big[ 1 - e^{- (\sqrt{p^2 + M^2} - H)/T} \Big] \Bigg\}  \nonumber \\
& = & \frac{1}{4 \pi T^3} \! \int_0^{\infty} \!\! \mbox{d}p \, p^3 \, \frac{1}{\sqrt{p^2 + M^2}} \,
\Bigg\{ \frac{1}{e^{(\sqrt{p^2 + M^2} + H)/T} - 1} + \frac{1}{e^{(\sqrt{p^2 + M^2} - H)/T} - 1} \Bigg\} \, , \nonumber
\end{eqnarray}
and
\begin{eqnarray}
\label{defh1}
{\hat h_1} & = & \frac{{\hat g}_1}{T} \\
& = & \frac{1}{4 \pi T} \! \int_0^{\infty} \!\! \mbox{d}p \, p \, \frac{1}{\sqrt{p^2 + M^2}} \,  \Bigg\{ \frac{1}{e^{(\sqrt{p^2 + M^2} + H)/T} - 1}
+ \frac{1}{e^{(\sqrt{p^2 + M^2} - H)/T} - 1}  \Bigg\} \, , \nonumber
\end{eqnarray}
respectively.

We are now able to extract the genuine spin-wave interaction part $z^{int}$ in the free energy density that is given by the difference
between the total two-loop free energy density $z$, derived in Ref.~\citep{Hof20a},
\begin{eqnarray}
\label{freeEDtwoLoopParallel}
z & = & z_0 - {\hat g}_0 + \frac{H}{\rho_s} \, {\hat g}_1 \, \frac{\partial {\hat g}_0}{\partial H}
- \frac{\sqrt{M_s H_s} H}{4 \pi \rho_s^{3/2}} \, \frac{\partial {\hat g}_0}{\partial H}
- \frac{H^2}{\rho_s}{( {\hat g}_1)}^2
+ \frac{\sqrt{M_s H_s} H^2}{2 \pi \rho_s^{3/2}} \, {\hat g}_1 \, , \nonumber \\
& & z_0 = - M_s H_s - \frac{M^{3/2}_s H^{3/2}_s}{6 \pi \rho_s^{3/2}} - (k_2 + k_3) \frac{M^2_s H^2_s}{\rho_s^2}
- \frac{M_s H_s  H^2}{16 \pi^2 \rho_s^2} \, ,
\end{eqnarray}
and the piece $z^{free}$, Eq.~(\ref{freeDressedED}), as
\begin{equation}
z^{int} = z - z^{free} \, .
\end{equation}
We obtain the simple result
\begin{eqnarray}
\label{fedTwoLoopDRESSED}
z^{int} & = & \frac{H}{\rho_s} \,{\hat h}_1 \, \frac{\partial {\hat h}_0}{\partial H} \, T^4
- \frac{H^2}{\rho_s} \, {({\hat h}_1)}^2  \, T^2 + z^{int}_0 \nonumber \\
& & z^{int}_0 = - \frac{M_s H_s  H^2}{16 \pi^2 \rho_s^2} \, ,
\end{eqnarray}
and furthermore identify $z^{free}_0$ as
\begin{equation}
z^{free}_0 = - M_s H_s - \frac{M^{3/2}_s H^{3/2}_s}{6 \pi \rho_s^{3/2}} - (k_2 + k_3) \frac{M^2_s H^2_s}{\rho_s^2} \, .
\end{equation}
The vacuum energy density involves the next-to-leading order effective constants $k_2$ and $k_3$. It should be emphasized that in two
spatial dimensions these are only relevant at zero temperature.\footnote{Besides, numerically they are small. For the spin-$\frac{1}{2}$
square-lattice antiferromagnet the value of the relevant combination $k_2 + k_3$ is known from Monte Carlo simulations \citep{GHJNW09}:
$(k_2 + k_3)/v^2 = -0.0018 \, \rho_s^{-1} = -0.0102 \, J^{-1}$.} The finite-temperature physics of the system, up to two-loop order, is fully
described in terms of the leading order effective constants $\rho_s$ and $M_s$. The only difference between, e.g., square and honeycomb
lattice antiferromagnets consists in the concrete values of $\rho_s$ and $M_s$. For the spin-$\frac{1}{2}$ square-lattice antiferromagnet
they are (see, e.g.,  Ref.~\citep{GHJNW09})
\begin{equation}
\label{squareLEC}
\rho_s = 0.1808(4) J \, , \quad M_s = 0.30743(1) / a^2 \, , \quad v = 1.6585(10) J a \, ,
\end{equation}
for the spin-$\frac{1}{2}$ honeycomb-lattice antiferromagnet, according to Ref.~\citep{JKNW08}, we have
\begin{equation}
\label{honeyLEC}
\rho_s = 0.102(2) J \, , \quad {\tilde M_s} = 0.2688(3) \, , \quad v = 1.297(16) J a \, ,
\end{equation}
where
\begin{equation}
{\tilde M_s} = \frac{ 3 \sqrt{3}}{4} \, M_s \, a^2 \, .
\end{equation}
Note that we also quote the respective values for the spin-wave velocity $v$.

\subsection{Three-Dimensional Antiferromagnets}
\label{dressed3d}

\begin{figure}
\begin{center}
\hbox{
\includegraphics[width=15.0cm]{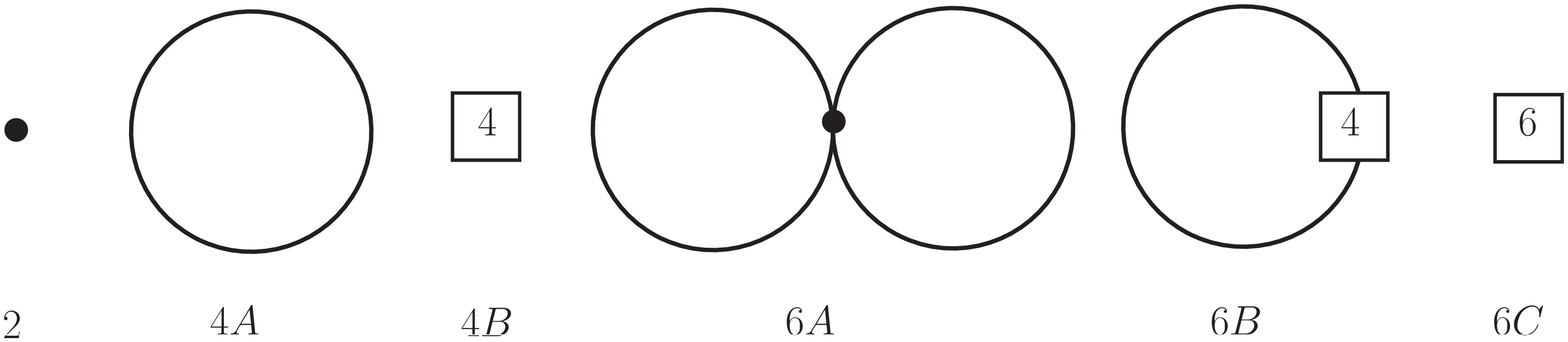}}
\vspace{4mm}
\hbox{
\includegraphics[width=12.6cm]{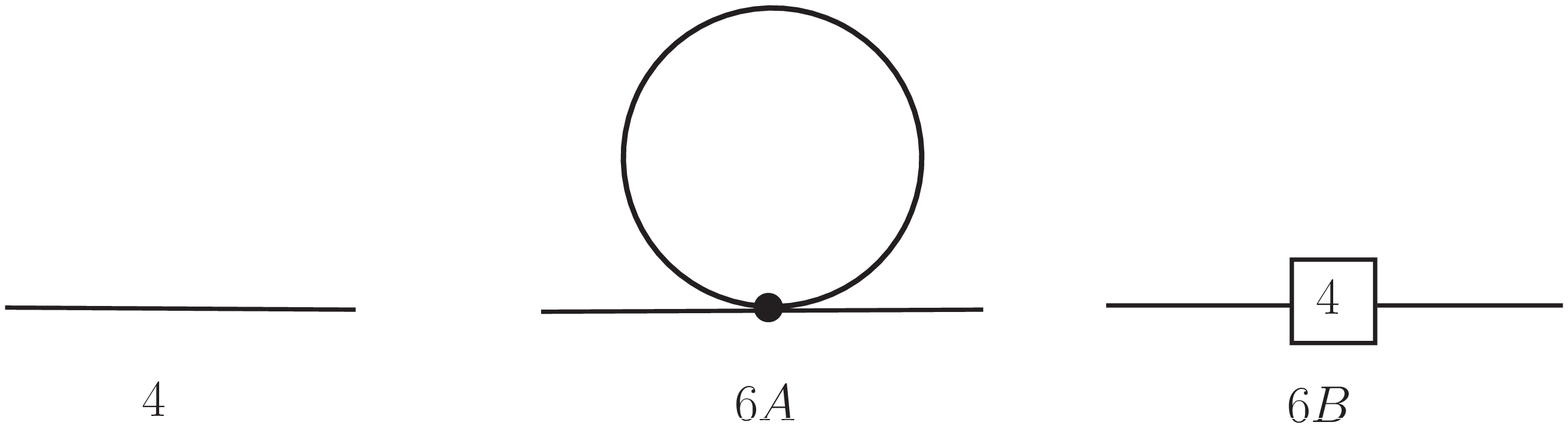}}
\end{center}
\caption{Three-dimensional antiferromagnets in mutually aligned magnetic and staggered fields. Upper panel: Partition function diagrams up
to two-loop order. Lower panel: Two-point function diagrams up to one-loop order. Filled circles constitute vertices from the leading order
effective Lagrangian, while the "4" ("6") in the box represents the NLO (NNLO) effective Lagrangian.}
\label{figure2}
\end{figure}

In three spatial dimensions, each additional magnon loop corresponds to a suppression of {\it two} powers of temperature. The diagrams for
the free energy density we need to evaluate up to two-loop order, are depicted in the upper panel of Fig.~\ref{figure2}. Comparing with the
relevant diagrams for two-dimensional antiferromagnets, we note that here two additional diagrams emerge: the one-loop graph $6B$ and the
tree graph $6C$. The leading finite-temperature contribution (order $T^4$) stems from the one-loop graph $4A$. At next-to-leading order we
have two finite-temperature contributions (order $T^6$) coming from the two-loop graph 6A and the one-loop graph 6B.\footnote{Details on
the effective loop evaluation can be found in Ref.~\citep{Hof20c}.}

To isolate the genuine magnon-magnon interaction portion in the free energy density, we follow the same strategy as before: we represent
the free energy density in terms of the dressed quasiparticles by evaluating the two-point function and the dispersion relation for both
antiferromagnetic magnons. The relevant diagrams for the two-point function $\tau_{\pm}(x-y)$ are shown in the lower panel of
Fig.~\ref{figure2}.

The leading contribution (diagram 4) corresponds to the zero-temperature propagator $\Delta_{\pm}(x-y)$ for magnon $+$ and magnon $-$,
\begin{equation}
\tau_{\pm}^4(x-y) = \Delta_{\pm}(x-y) = \int \!\! \frac{\mbox{d}^d \! p}{{(2 \pi)}^d} \,
\frac{e^{ip(x-y)}}{p_4^2 + {\vec p \,}^2 + M^2 \pm 2 i H p_4 - H^2} \, .
\end{equation}
The correction from the one-loop graph 6A is the same as for the quantities $\tau_{\pm}^{4a}(x-y)$, Eq.~(\ref{TwoPointA}),
\begin{equation}
\label{TwoPoint6a}
\tau_{\pm}^{6A}(x-y) = \pm \frac{2 i H}{\rho_s} \, \Delta(0) \, \int \!\! \frac{\mbox{d}^d \! p}{{(2 \pi)}^d}
\, \frac{p_4 \, e^{ip(x-y)}}{{(p_4^2 + {\vec p \,}^2 + M^2 \pm 2 i H p_4 - H^2)}^2} \, ,
\end{equation}
with the exception that the dimensionally regularized expression $\Delta(0)$ diverges when the physical limit $d\to4$ is taken. This
apparent dilemma is solved by the observation that the additional contribution from the tree graph $6B$,
\begin{eqnarray}
\label{TwoPoint6b}
\tau_{\pm}^{6B}(x-y) & = &  -\frac{2(k_2 - k_1) M^4}{\rho_s} \,
\int \!\! \frac{\mbox{d}^d \! p}{{(2 \pi)}^d} \, \frac{e^{ip(x-y)}}{{(p_4^2 + {\vec p \,}^2 + M^2 \pm 2 i H p_4 - H^2)}^2} \nonumber \\
& & \pm \frac{4 i k_1 M^2 H}{\rho_s} \,
\int \!\! \frac{\mbox{d}^d \! p}{{(2 \pi)}^d} \, \frac{p_4 \, e^{ip(x-y)}}{{(p_4^2 + {\vec p \,}^2 + M^2 \pm 2 i H p_4 - H^2)}^2} \, ,
\end{eqnarray}
also becomes singular in the limit $d\to4$ and that the divergences mutually cancel as we now show.

The divergence in the zero-temperature propagator at the coordinate origin, $\Delta(0)$,
\begin{equation}
\label{regprop2}
\Delta(0) = \int \frac{{\mbox{d}}^d p}{{(2 \pi)}^d} \, \frac{1}{p^2 + M^2}
= {\int}_{\!\!\!0}^{\infty} \mbox{d} \rho \, (4 \pi \rho)^{-d/2} e^{-\rho M^2}
= 2 M^2 \lambda \, ,
\end{equation}
is contained in the parameter $\lambda$,
\begin{eqnarray}
\label{lambda}
\lambda & = & \mbox{$ \frac{1}{2}$} \, (4 \pi)^{-d/2} \, \Gamma(1-{\mbox{$ \frac{1}{2}$}}d) M^{d-4} \nonumber\\
& = & \frac{M^{d-4}}{16{\pi}^2} \, \Bigg[ \frac{1}{d-4} - \mbox{$ \frac{1}{2}$} \{ \ln{4{\pi}} + {\Gamma}'(1) + 1 \}
+ {\cal O}(d-4) \Bigg] \, .
\end{eqnarray}
On the other hand, the singularity in diagram $6B$ is due to the presence of NLO effective constants. Following Ref.~\cite{Hof17b,Hof20c},
these can be written as
\begin{equation}
k_1 = {\tilde \gamma}_3 \Big( \lambda + \frac{{\overline k}_1}{32 \pi^2} \Big) \, , \quad
k_2 = {\tilde \gamma}_4 \Big( \lambda + \frac{{\overline k}_2}{32 \pi^2} \Big) \, ,
\end{equation}
with coefficients
\begin{equation}
{\tilde \gamma}_3 = -1 \, , \quad
{\tilde \gamma}_4 = -1 \, .
\end{equation}
The  $\lambda$-divergences in the sum of diagrams $6A$ and $6B$ hence cancel. The renormalized NLO effective constants ${\overline k}_1$
and ${\overline k}_2$ are finite in $d=4$ and of unit order, 
\begin{equation}
{\overline k}_1, {\overline k}_2 \approx 1 \, .
\end{equation} 

Adhering to the same steps as in the previous subsection, the magnon dispersion relations take the form
\begin{eqnarray}
\omega_{\pm} & = & \sqrt{{\vec p \,}^2 + \frac{M_s H_s}{\rho_s} - \frac{{\overline k}_2 - {\overline k}_1}{16 \pi^2 \rho_s} \,
{\Bigg( \frac{M_s H_s}{\rho_s} \Bigg)}^2
+ \frac{{\overline k}_1}{8 \pi^2 \rho_s} \, \frac{M_s H_s H^2}{\rho_s}} \nonumber \\
& & \pm H \pm \frac{{\overline k}_1}{16 \pi^2 \rho_s} \, \frac{M_s H_s H}{\rho_s} \, ,
\end{eqnarray}
and the piece in the free energy density that is due to noninteracting magnons amounts to
\begin{eqnarray}
\label{freeDressedED3d}
z^{free} & = & - {\hat h}_0 \, T^4
- \frac{{\overline k}_2 - {\overline k}_1}{16 \pi^2} \frac{M^2_s H^2_s}{\rho^3_s} \, {\hat h}_1 T^2
- \frac{{\overline k}_1}{16 \pi^2} \, \frac{H M_s H_s}{\rho^2_s} \frac{\partial {\hat h}_0}{\partial H} \, T^4 \nonumber \\
& & + \frac{{\overline k}_1}{8 \pi^2} \, \frac{H^2 M_s H_s}{\rho^2_s} \, {\hat h}_1 T^2
+ z^{free}_0 \, .
\end{eqnarray}
The kinematical functions in three spatial dimensions, ${\hat h}_0$ (or equivalently: ${\hat g}_0)$ and ${\hat h}_1$ (or equivalently:
${\hat g}_1)$, are  
\begin{eqnarray}
\label{defh0d3}
{\hat h}_0 & = &  \frac{{\hat g}_0}{T^4} \\
& = & - \frac{1}{2 \pi^2 T^3} \int_0^{\infty} \!\! \mbox{d}p \, p^2 \, \Bigg\{ \ln \Big[ 1 - e^{- (\sqrt{p^2 + M^2} + H)/T} \Big] +
\ln \Big[ 1 - e^{- (\sqrt{p^2 + M^2} - H)/T} \Big] \Bigg\} \nonumber \\
& = & \frac{1}{6 \pi^2 T^4} \! \int_0^{\infty} \!\! \mbox{d}p \, p^4 \, \frac{1}{\sqrt{p^2 + M^2}} \,
\Bigg\{ \frac{1}{e^{(\sqrt{p^2 + M^2} + H)/T} - 1} +  \frac{1}{e^{(\sqrt{p^2 + M^2} - H)/T} - 1} \Bigg\} \, , \nonumber
\end{eqnarray}
and
\begin{eqnarray}
\label{defh1d3}
{\hat h_1} & = &  \frac{{\hat g}_1}{T^2} \\
& = & \frac{1}{4 \pi^2 T^2} \! \int_0^{\infty} \!\! \mbox{d}p \, p^2 \, \frac{1}{\sqrt{p^2 + M^2}} \,
\Bigg\{ \frac{1}{e^{(\sqrt{p^2 + M^2} + H)/T}- 1} + \frac{1}{e^{(\sqrt{p^2 + M^2} - H)/T} - 1} \Bigg\} \, , \nonumber
\end{eqnarray}
respectively.

Resorting to the renormalized representation for the total two-loop free energy density, derived in Ref.~\citep{Hof20c},
\begin{eqnarray}
\label{freeED2dtwoLoopParallel}
z & = & z_0 - {\hat g}_0 + \frac{H}{\rho_s} \, {\hat g}_1 \, \frac{\partial {\hat g}_0}{\partial H}
- \frac{H^2}{\rho_s}{( {\hat g}_1)}^2
- \frac{{\overline k}_2 - {\overline k}_1}{16 \pi^2} \frac{M^2_s H^2_s}{\rho^3_s} \, {\hat g}_1
- \frac{{\overline k}_1}{16 \pi^2} \, \frac{H M_s H_s}{\rho^2_s} \frac{\partial {\hat g}_0}{\partial H} \nonumber \\
& & + \frac{{\overline k}_1}{8 \pi^2} \, \frac{H^2 M_s H_s}{\rho^2_s} \, {\hat g}_1 \, , \nonumber \\
& & z_0 = - M_s H_s + \frac{ {\overline k}_2 - 2{\overline k}_3}{32 \pi^2} \, \frac{M^2_s H^2_s}{\rho^2_s}
- \frac{M^2_s H^2_s}{64 \pi^2 \rho^2_s} + {\cal O}(p^6) \, ,
\end{eqnarray}
we extract the genuine spin-wave interaction part $z^{int}$ as
\begin{eqnarray}
\label{fedTwoLoopDRESSED3d}
z^{int} & = & \frac{H}{\rho_s} \,{\hat h}_1 \, \frac{\partial {\hat h}_0}{\partial H} \, T^6
- \frac{H^2}{\rho_s} \, {({\hat h}_1)}^2  \, T^4 + z^{int}_0 \nonumber \\
& & z^{int}_0 = {\cal O}(p^6) \, .
\end{eqnarray}
The renormalized NLO effective constant ${\overline k}_3$, much like ${\overline k}_1$ and ${\overline k}_2$, is of natural order, i.e.,
${\overline k}_3 \approx 1$.

Finally, we identify the vacuum energy density associated with noninteracting magnons as
\begin{equation}
\label{z0free}
z^{free}_0 = - M_s H_s + \frac{ {\overline k}_2 - 2{\overline k}_3}{32 \pi^2} \, \frac{M^2_s H^2_s}{\rho^2_s}
- \frac{M^2_s H^2_s}{64 \pi^2 \rho^2_s} \, .
\end{equation}
Unlike for antiferromagnetic films, in three spatial dimensions, NLO effective constants also show up in the finite-temperature piece
$z - z_0$ according to Eq.~(\ref{freeED2dtwoLoopParallel}). The remainder ${\cal O}(p^6)$ of the zero-temperature contribution $z_0$ even
contains NNLO effective constants that originate from the tree graph $6C$ of Fig.~\ref{figure2}. The corresponding terms in the vacuum
energy density are of the form $\propto H_s^3, \propto H_s^2 H^2, \propto H_s H^4$ and $\propto H^6$, where each such term contains a linear
combination of NNLO effective constants, much like the second term in Eq.~(\ref{z0free}) involves the combination
${\overline k}_2 - 2{\overline k}_3$ of NLO effective constants. But because the numerical values of NNLO effective constants are very
small -- and their sign a priori unknown -- we refrain from providing a lenghty explicit expression for all these higher-order
contributions that only matter at zero temperature.

It should be stressed that the finite-temperature interaction contribution $z^{int}-z^{int}_0$ is free of such NLO (or NNLO) quantities. As
in the case of antiferromagnetic films, it only depends on the spin stiffness $\rho_s$ and the order parameter $M_s$. This implies that the
question of whether the magnon-magnon interaction in the pressure leads to attraction or repulsion, can be answered in a model-independent
-- and thus universal -- way also in three spatial dimensions.

\section{Magnon Pressure and Interaction Effects}
\label{pressure}

In this section -- within the dressed magnon picture -- we provide the low-temperature representation for the pressure and study how it is
affected by mutually parallel magnetic and staggered fields. Of particular interest is the question whether the magnon-magnon interaction
causes an attraction or a repulsion in the pressure. Again, we treat two- and three-dimensional antiferromagnets separately.

\subsection{Two-Dimensional Antiferromagnets}
\label{pressure2d}

The thermodynamic quantities depend on three parameters: $T, H_s, H$. Magnon effective field theory is valid in the sector where these
quantities are small, i.e., small with respect to the exchange integral $J$ that defines the non-thermal microscopic scale in the
underlying Hamiltonian.

Rather than operating with the dimensionful quantities $T, H_s, H$, we define three dimensionless parameters as
\begin{equation}
\label{definitionRatios2d}
t \equiv \frac{T}{2 \pi \rho_s} \, , \qquad
m \equiv \frac{\sqrt{M_s H_s}}{2 \pi \rho_s^{3/2}} \, , \qquad
m_H \equiv \frac{H}{2 \pi \rho_s} \, .
\end{equation}
Note that the common denominator,
\begin{equation}
2 \pi \rho_s \approx J \, ,
\end{equation}
approximately concurs with the exchange coupling $J$, such that $t,m,m_H$ must be small. In concrete plots we choose the parameter region
as
\begin{equation}
\label{domain2d}
t, m, m_H \ \lesssim 0.4 \, .
\end{equation}
In addition, we implement the stability criterion, Eq.~(\ref{stabilityCondition}), by
\begin{equation}
m > m_H + \delta \, , \qquad  \delta = 0.1 \, .
\end{equation}

\begin{figure}
\begin{center}
\hbox{
\includegraphics[width=7.6cm]{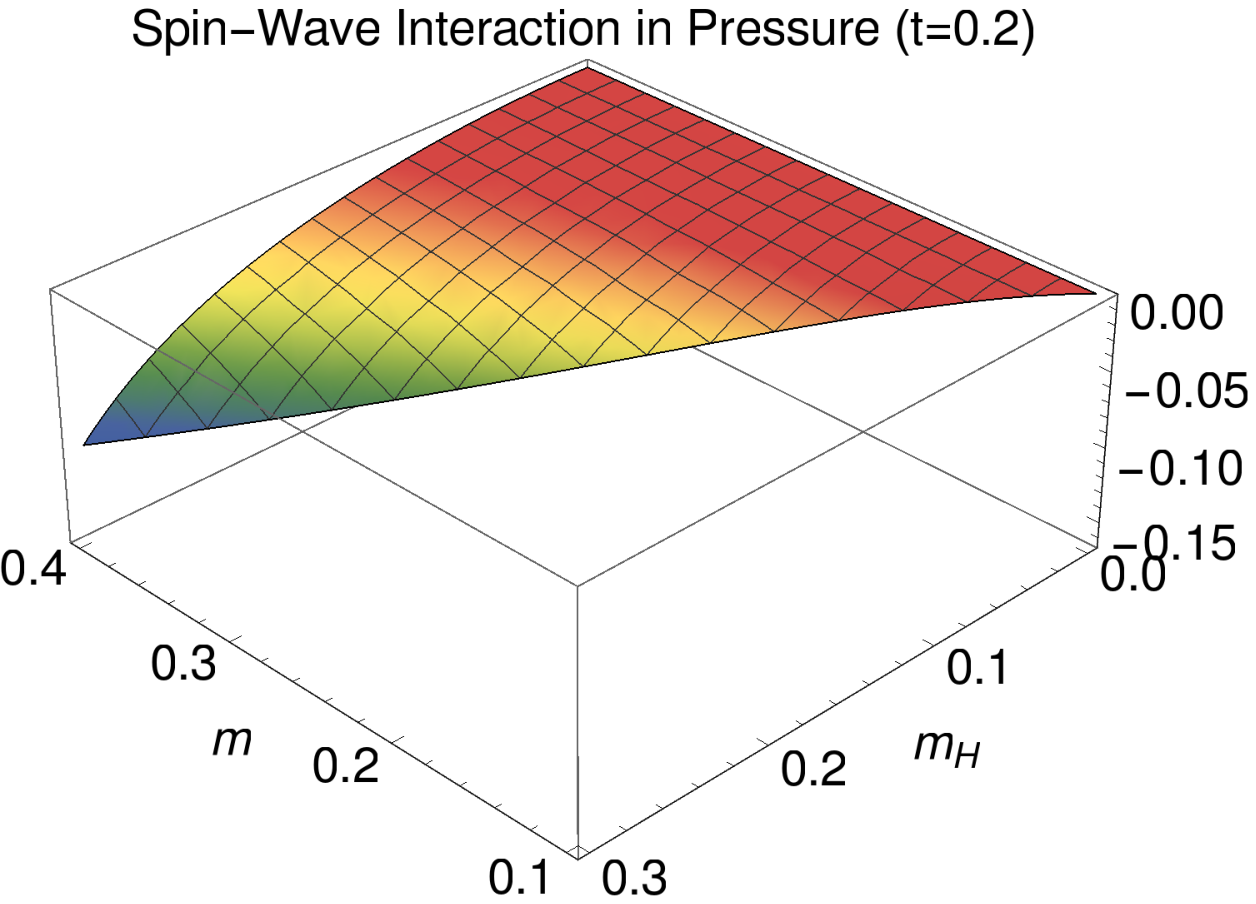}
\includegraphics[width=7.6cm]{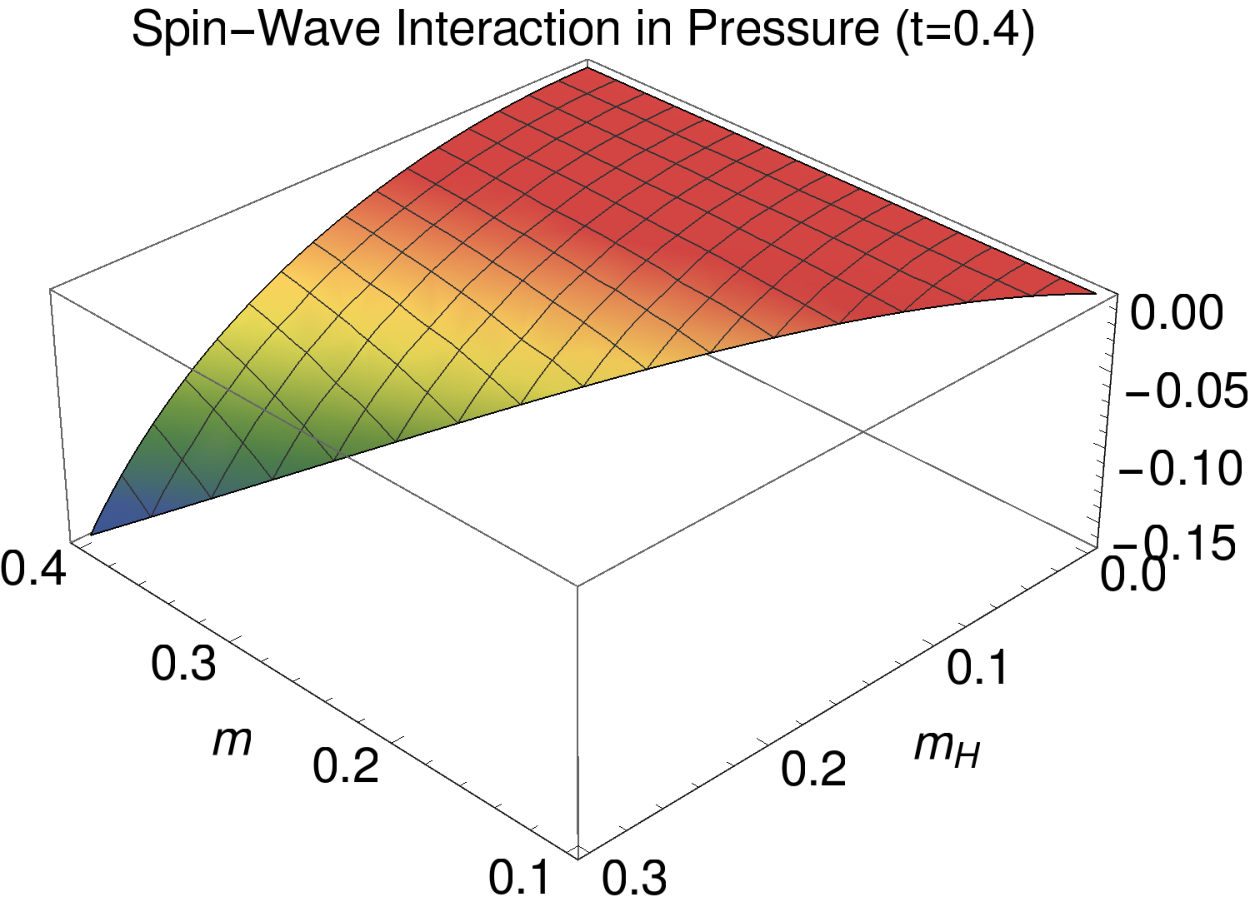}}
\end{center}
\caption{[Color online] $P_{int}$: Impact of the genuine spin-wave interaction on the pressure of a bipartite two-dimensional
antiferromagnet in mutually aligned magnetic ($m_H$) and staggered ($m$) fields at the temperatures $t = 0.2$ and $t = 0.4$.}
\label{figure3}
\end{figure}

The low-temperature series for the pressure, i.e., the negative of the temperature-dependent part of the free energy density,
\begin{equation}
\label{defPressure}
P = z_0 - z \, ,
\end{equation}
takes the form
\begin{equation}
\label{pressureAF}
P(t,m,m_H) = p_1 T^3 + p_2 T^4 + {\cal O}(T^5) \, .
\end{equation}
The coefficient $p_1$ of the dominant piece ($\propto T^3$) stems from noninteracting dressed magnons and reads
\begin{equation}
p_1 = {\hat h}_0 + \frac{m m_H}{2} \, \frac{\partial {\hat h}_0}{\partial m_H} - \frac{m m_H^2}{t^2} \, {\hat h}_1 \, ,
\end{equation}
while interaction effects are contained in the order-$T^4$ contribution with coefficient
\begin{equation}
\label{pressureAFfree}
p_2 = \Bigg\{ - 2 \pi m_H t \, {\hat h}_1 \frac{\partial {\hat h}_0}{\partial m_H} + \frac{2 \pi m_H^2}{t} \, {({\hat h}_1)}^2
\Bigg\} \, \frac{1}{2 \pi \rho_s t} \, .
\end{equation}

To explore the effect of the magnon-magnon interaction in the pressure, we define the dimensionless ratio between interaction piece and
free dressed magnon gas as
\begin{equation}
\label{intRatioP}
P_{int} = \frac{p_2 T^4}{p_1 T^3} = 2 \pi \rho_s t \, \frac{p_2}{p_1} \, .
\end{equation}
In Fig.~\ref{figure3}, for a bipartite two-dimensional antiferromagnet at the temperatures $t = 0.2$ and $t = 0.4$, the ratio $P_{int}$ is
plotted as a function of magnetic ($m_H$) and staggered ($m$) field strength.\footnote{Note that the spin stiffness $\rho_s$ drops out in
the ratio $P_{int}$: we are hence dealing with a universal parametrization valid for a generic two-dimensional bipartite antiferromagnet.}
One observes that the interaction in stronger fields is quite large, amounting up to about fifteen percent compared to the noninteracting
magnon gas contribution. In the entire parameter space we consider, the genuine spin-wave interaction in the pressure is attractive and
tends to zero when the magnetic field is turned off.

\begin{figure}
\begin{center}
\hbox{
\includegraphics[width=7.8cm]{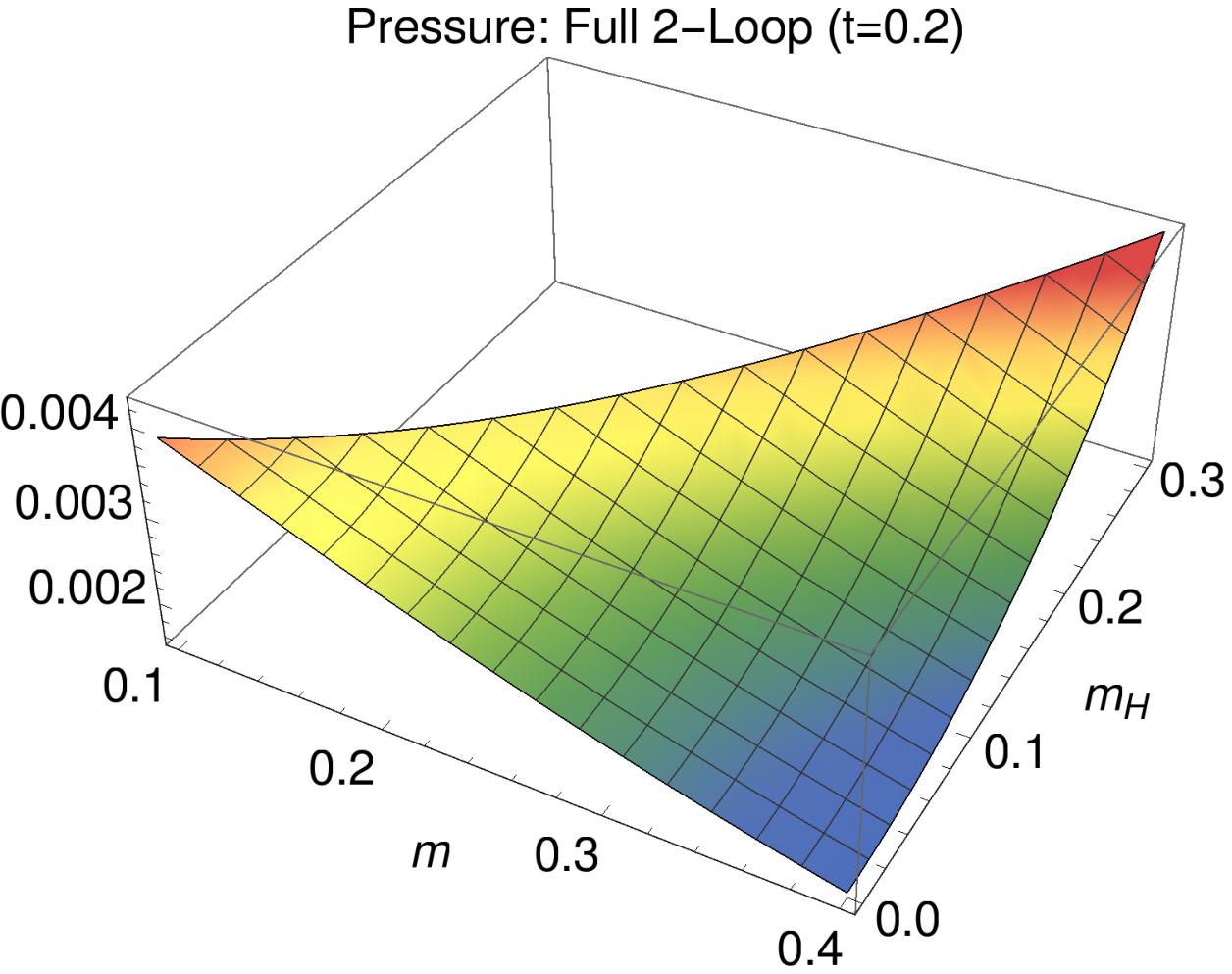}
\includegraphics[width=7.8cm]{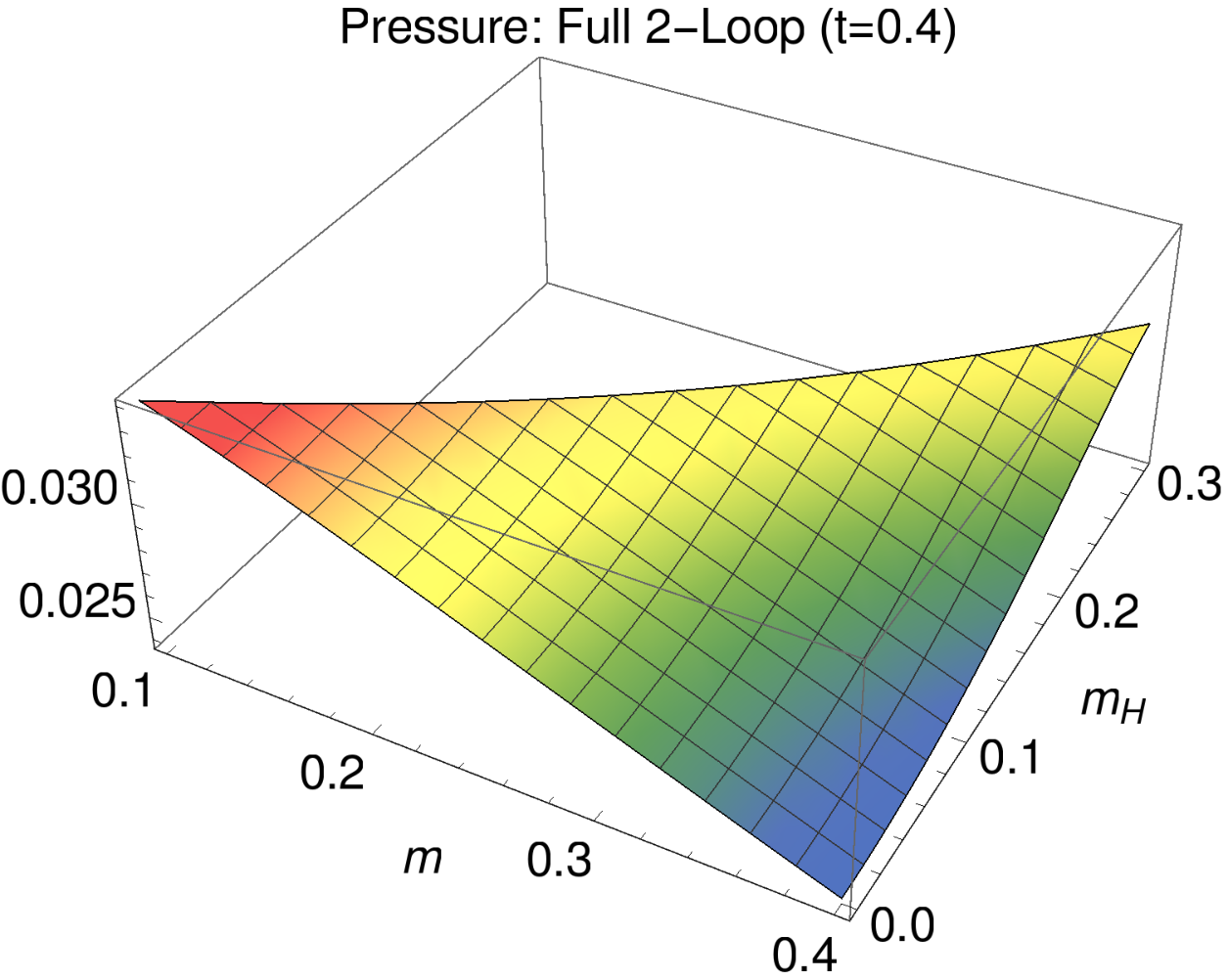}}
\end{center}
\caption{[Color online] Total two-loop representation for pressure -- $p_1 T^3 + p_2 T^4$ -- for the spin-$\frac{1}{2}$ square-lattice
antiferromagnet in mutually aligned magnetic ($m_H$) and staggered ($m$) fields at the temperatures $t = 0.2$ and $t = 0.4$.}
\label{figure4}
\end{figure}

In Fig.~\ref{figure4}, for the same temperatures $t = 0.2$ and $t = 0.4$, we show the full two-loop representation for the pressure, i.e.,
the quantity 
\begin{equation}
p_1 T^3 + p_2 T^4 \, ,
\end{equation}
specifically for the spin-$\frac{1}{2}$ square-lattice antiferromagnet. One identifies two opposite tendencies: the pressure grows as the
magnetic field gets stronger, but the pressure drops when the staggered field increases. It should be noted that the pressure -- up to
two-loop order -- does not involve any microscopic quantities other than $\rho_s$ and $M_s$.\footnote{$M_s$ is hidden in the low-energy
parameter $m$.}

\subsection{Three-Dimensional Antiferromagnets}
\label{pressure3d}

To define analogous low-energy parameters $t,m,m_H$ for antiferromagnets in three spatial dimensions, we consider the simple cubic
spin-$\frac{1}{2}$ antiferromagnet, where spin stiffness and exchange integral are connected by (see Ref.~\citep{Hof99b})
\begin{equation}
\label{spinStiffnessd3}
\sqrt{\rho_s} \approx 0.61 |J| \, .
\end{equation}
Accordingly, the three dimensionless parameters we define as
\begin{equation}
\label{definitionRatios}
t \equiv \frac{T}{\sqrt{\rho_s}} \, , \qquad
m \equiv \frac{\sqrt{M_s H_s}}{\rho_s} \, , \qquad
m_H \equiv \frac{H}{\sqrt{\rho_s}} \, .
\end{equation}
Here we choose the parameter range as
\begin{equation}
\label{domain}
t, \, m, \, m_H \ \lesssim 0.6 \, ,
\end{equation}
and implement the stability criterion, Eq.~(\ref{stabilityCondition}), again by
\begin{equation}
m > m_H + \delta \, , \qquad  \delta = 0.1 \, .
\end{equation}

The low-temperature expansion of the pressure for three-dimensional antiferromagnets takes the structure
\begin{equation}
P(t,m,m_H) =  p_1 T^4 + p_2 T^6 + {\cal O}(T^8) \, .
\end{equation}
The coefficient $p_1$ of the dominant piece refers to noninteracting dressed magnons,
\begin{equation}
p_1 = {\hat h}_0 + \frac{{\overline k}_2 - {\overline k}_1}{16 \pi^2} \, \frac{m^4}{t^2} \, {\hat h}_1
+ \frac{{\overline k}_1}{16 \pi^2} \, m^2 m_H \frac{\partial {\hat h}_0}{\partial m_H}
- \frac{{\overline k}_1}{8 \pi^2} \, \frac{m^2 m^2_H }{t^2}\, {\hat h}_1 \, ,
\end{equation}
while the subsequent contribution of order $T^6$,
\begin{equation}
\label{freeEnergyDensity}
p_2 = \frac{1}{\rho_s t^2} \, \Bigg\{ -m_H t^2 \, {\hat h}_1 \, \frac{\partial {\hat h}_0}{\partial m_H}
+ m^2_H \, {({\hat h}_1 )}^2 \Bigg\} \, ,
\end{equation}
corresponds to the magnon-magnon interaction.

\begin{figure}
\begin{center}
\hbox{
\includegraphics[width=7.6cm]{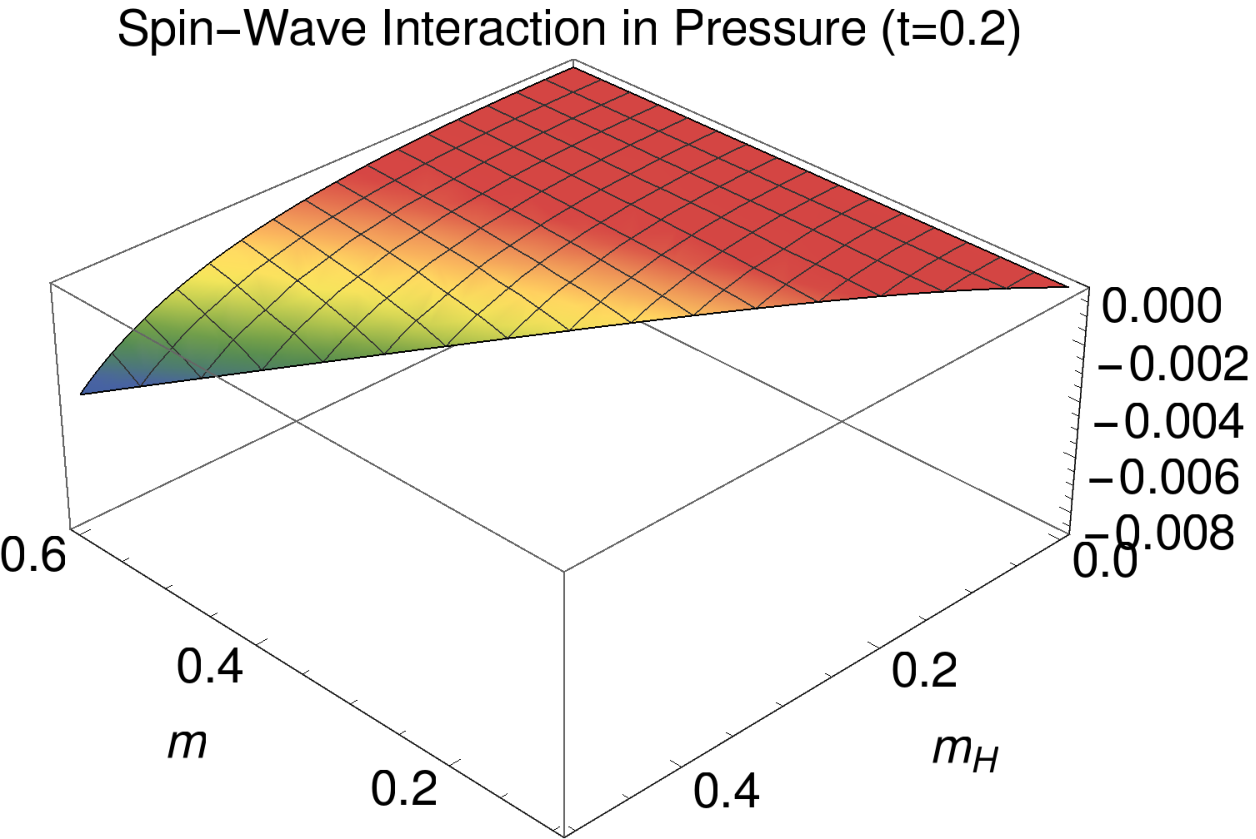} 
\includegraphics[width=7.6cm]{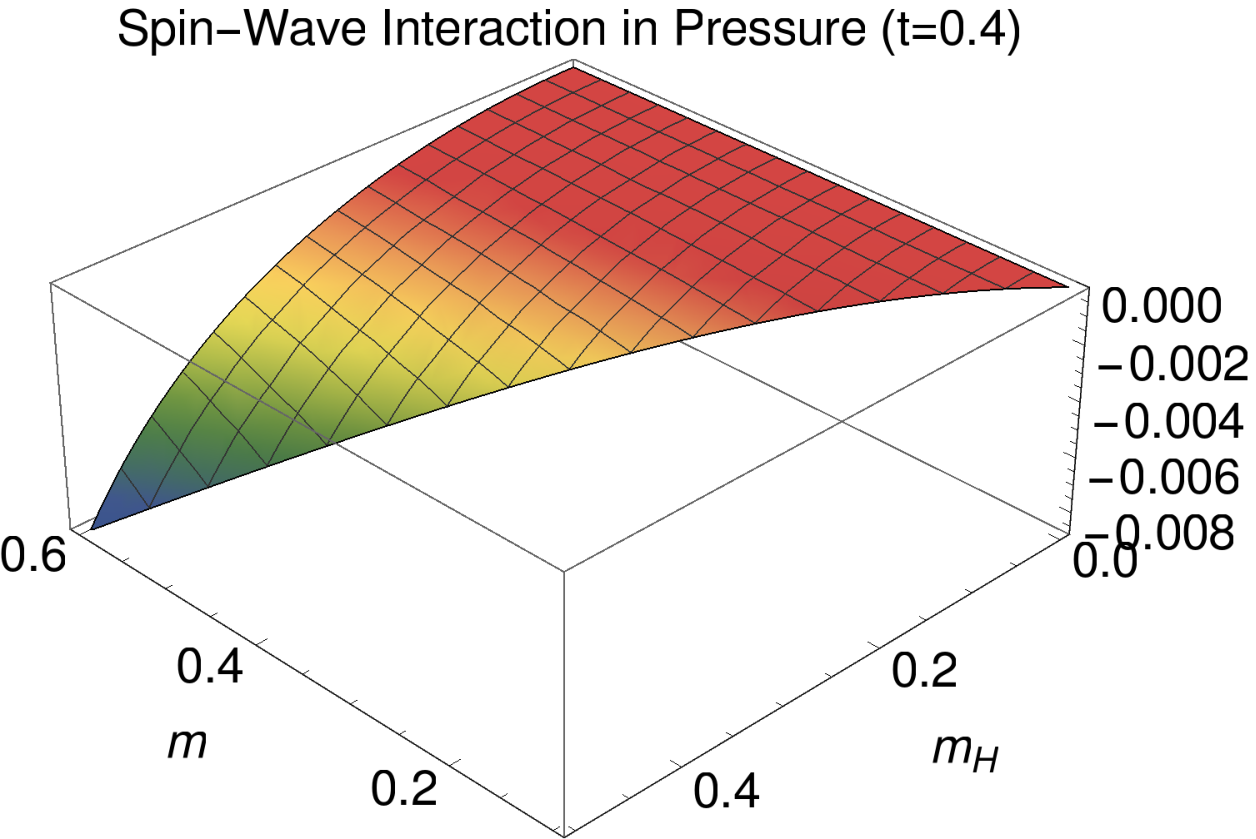}}
\end{center}
\caption{[Color online] $P_{int}$: Impact of the genuine spin-wave interaction on the pressure of a bipartite three-dimensional
antiferromagnet in mutually aligned magnetic ($m_H$) and staggered ($m$) fields at the temperatures $t = 0.2$ and $t = 0.4$.}
\label{figure5}
\end{figure}

The leading coefficient $p_1$ is dominated by the kinematical function ${\hat h}_0$: the remaining terms that contain the NLO effective
constants ${\overline k}_1$ and ${\overline k}_2$ are small. Since only order of magnitude of ${\overline k}_1$ and ${\overline k}_2$ is
known -- but not their exact numerical values for concrete physical samples -- in our assessment of the magnon-magnon interaction we
consider the dimensionless ratio
\begin{equation}
\label{intRatioPd3}
P_{int} = \frac{p_2 T^6}{{\hat h}_0 T^4} = \rho_s t^2 \, \frac{p_2}{{\hat h}_0} \, .
\end{equation}

In Fig.~\ref{figure5}, for a generic bipartite three-dimensional antiferromagnet at the temperatures $t = 0.2$ and $t = 0.4$, the ratio
$P_{int}$ is plotted as a function of magnetic ($m_H$) and staggered ($m$) field strength. In contrast to the two-dimensional case, here the
interaction is rather weak: even in stronger fields it only amounts up to about one percent compared to the noninteracting magnon gas. But
we find that the genuine spin-wave interaction in the pressure is attractive also in the case of three-dimensional antiferromagnets and
that it tends to zero when the magnetic field is turned off.

\begin{figure}
\begin{center}
\hbox{
\includegraphics[width=7.8cm]{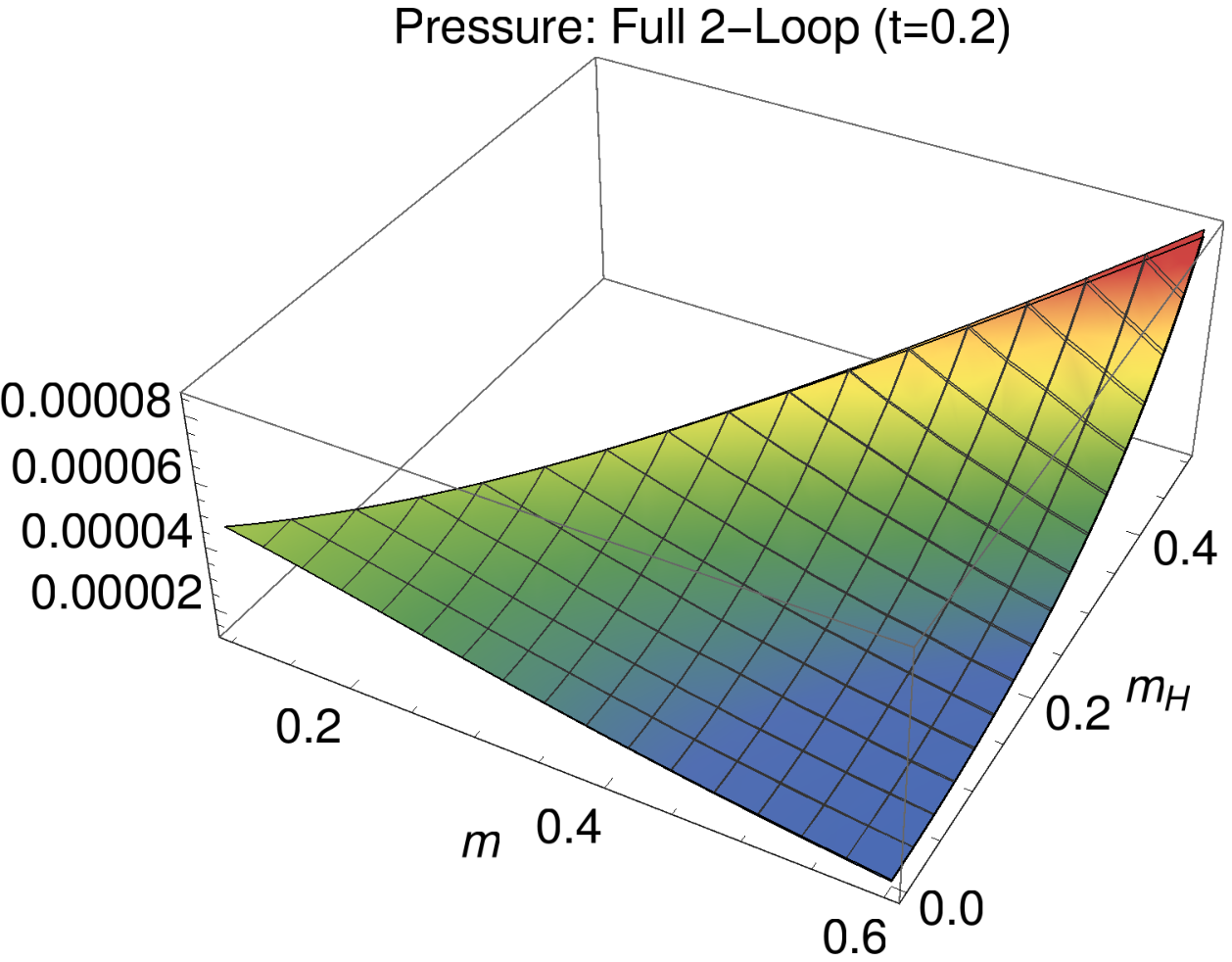}
\includegraphics[width=7.8cm]{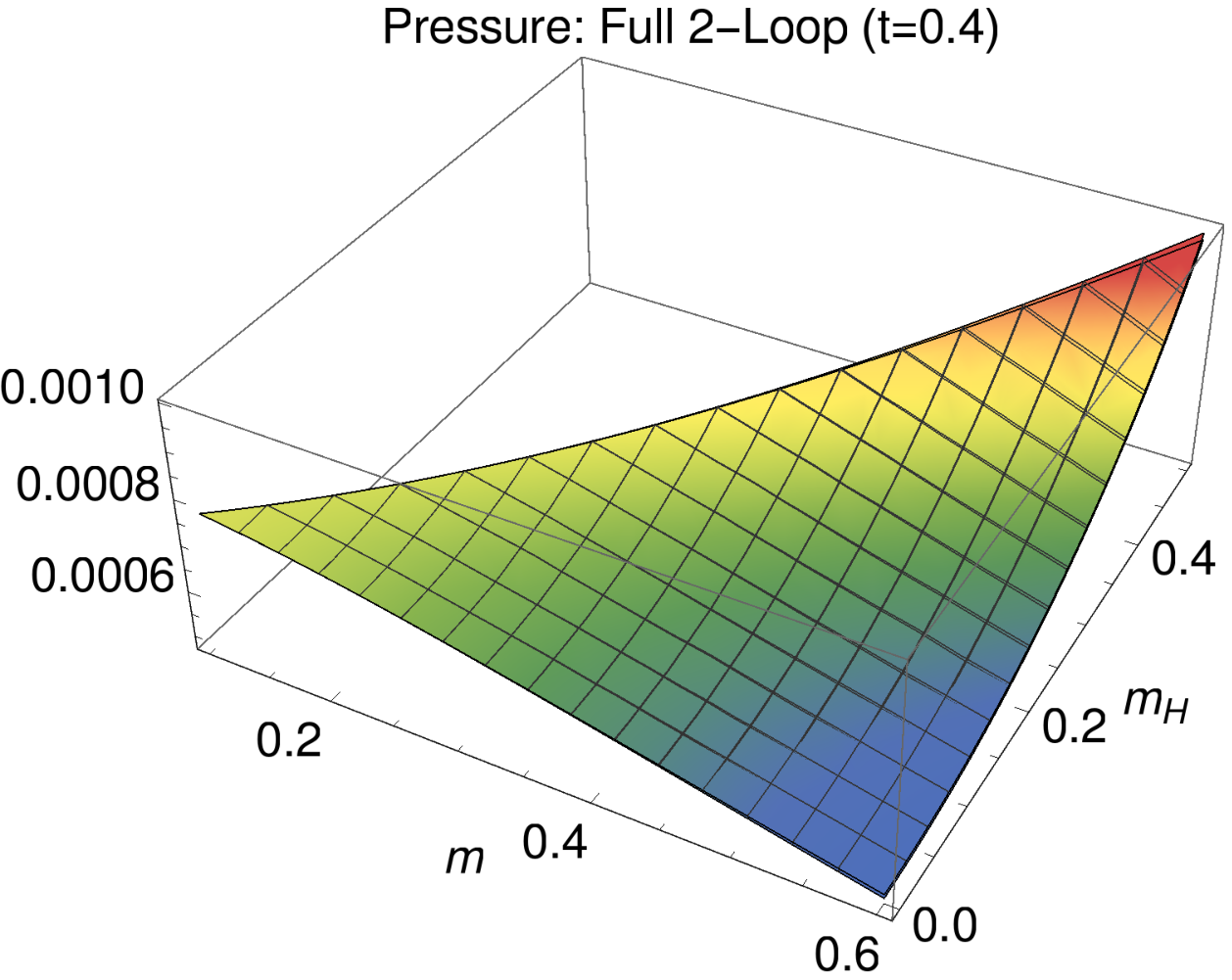}}
\end{center}
\caption{[Color online] Total two-loop representation for pressure -- $p_1 T^4 + p_2 T^6$ -- for the simple cubic spin-$\frac{1}{2}$
antiferromagnet in mutually aligned magnetic ($m_H$) and staggered ($m$) fields at the temperatures $t = 0.2$ and $t = 0.4$.}
\label{figure6}
\end{figure}

For the same temperatures $t = 0.2$ and $t = 0.4$, in Fig.~\ref{figure6}, we show the full two-loop representation for the pressure, i.e.,
the quantity 
\begin{equation}
p_1 T^4 + p_2 T^6
\end{equation}
for the simple cubic spin-$\frac{1}{2}$ antiferromagnet. As before we identify two opposite tendencies: the pressure grows when the
magnetic field gets stronger, but the pressure drops as the staggered field increases.

One final comment is in order here. Fig.~\ref{figure6} refers to the simple cubic spin-$\frac{1}{2}$ antiferromagnet where the numerical
value of the spin stiffness, Eq.~(\ref{spinStiffnessd3}), is known. However, we are unaware of the precise values of the NLO effective
constants ${\overline k}_1$ and ${\overline k}_2$. But since we know that their magnitude is of order one, we can perform a scan of these
quantities in the interval
\begin{equation}
\label{scanningk1k2}
\{ {\overline k}_1, {\overline k}_2 \} \ \subset \ [ -5, 5 ] \, ,
\end{equation}
which gives us a set of surfaces for the pressure $P(t,m,m_H)$. From these scans we then select the respective two extreme situations:
maximal and minimal corrections for each point in parameter space $(t,m,m_H)$. These two surfaces represent estimates of upper and lower
bounds for the corrections that are due to ${\overline k}_1$ and ${\overline k}_2$. As witnessed by Fig.~\ref{figure6}, these two extreme
hypersurfaces can barely be distinguished even in stronger fields, i.e., the corrections involving NLO effective constants are indeed very
small.

\section{Conclusions}
\label{conclusions}

In the first part of our systematic effective field theory investigation of antiferromagnetic films and solids in mutually parallel
magnetic and staggered fields, we derived the two-point function up to one-loop order and obtained corrections in the dispersion relation
for the two magnons. On the basis of these results we could extract the genuine magnon-magnon interaction piece in the two-loop free energy
density.

Within this dressed magnon picture we then showed that the interaction in the pressure is attractive both for two and three-dimensional
bipartite antiferromagnets -- but quite small in the latter case. While concrete plots for the full two-loop representation of the pressure
referred to the spin-$\frac{1}{2}$ square-lattice and the simple cubic spin-$\frac{1}{2}$ antiferromagnet, our results are fully predictive
for arbitrary bipartite geometry.

Moreover, the genuine magnon-magnon interaction portion in the pressure does not involve -- neither in two nor in three spatial dimensions
-- any NLO effective constants, but is completely fixed by the spin stiffness and the order parameter, i.e., the staggered magnetization at
zero temperature. In this sense, the outcome that the interaction in the pressure is attractive can be considered as universal.

\end{document}